\documentclass[onecolumn]{aa}
\usepackage{psfig}
\def\simlt{\ \raise -2.truept\hbox{\rlap{\hbox{$\sim$}}\raise5.truept   %
\hbox{$<$}\ }}
\def\simgt{\ \raise -2.truept\hbox{\rlap{\hbox{$\sim$}}\raise5.truept   %
\hbox{$>$}\ }}                                                          %

\def\be{\begin{equation}}
\def\ee{\end{equation}}
\def\newline{\hfil\break}

\def\ergs{{erg~cm$^{-2}$s$^{-1}$~}}

\def\la{\mathrel{\hbox{\rlap{\hbox{\lower4pt\hbox{$\sim$}}}\hbox{$<$}}}}
\def\ga{\mathrel{\hbox{\rlap{\hbox{\lower4pt\hbox{$\sim$}}}\hbox{$>$}}}}

\begin{document}

\title{Evidence for a Significant Blazar Contamination in CMB Anisotropy Maps}

   \author{P. Giommi \inst{1} and S. Colafrancesco \inst{2}}

   \offprints{P. Giommi}

\institute{ ASI Science Data Center, ASDC, Agenzia Spaziale Italiana c/o ESRIN,
            via G. Galilei 00044 Frascati, Italy.
            Email: paolo.giommi@asi.it
\and
              INAF - Osservatorio Astronomico di Roma
              via Frascati 33, I-00040 Monteporzio, Italy.
              Email: cola@mporzio.astro.it
             }

\date{Received / Accepted }
\authorrunning {P. Giommi \& S. Colafrancesco}
\titlerunning {Blazars in CMB maps}

\abstract{The analysis of the recent WMAP source catalog shows that the vast
majority of bright foreground extragalactic sources detected in CMB maps are
Blazars. In this paper we calculate the contamination of CMB anisotropy maps by
this type of flat-spectrum, strongly variable and polarized extragalactic radio
sources using up-to-date results from recent deep multi-frequency surveys.
From a careful archive search and from multi-frequency catalog
cross-correlations we found that more than 50 known Blazars (or Blazar
candidates) expected to be above the sensitivity limit of the BOOMERANG
experiment are included in the 90/150 GHz BOOMERANG anisotropy maps, a factor
$\simgt 15$ larger than previously reported. Using a recent derivation of the
Blazar radio LogN-LogS we show that Blazars, whose counts
continue to grow steeply down to faint fluxes, can sensitively affect CMB
fluctuation maps with even moderate resolution and sensitivity. We calculate
specifically that these sources induce an average sky brightness of $0.2
Jy/deg^2$, corresponding to an average temperature of $\approx 3-5 ~\mu$K.
Moreover, we find that the associated level of fluctuations is of the order of
$C_{\ell, Blazar}= 1.3\times 10^{-2}~\mu K^2 sr$ at 41 GHz and $C_{\ell,
Blazar}= 6.5\times 10^{-4}~\mu K^2 sr$ at 94 GHz.
Taking into account both Blazar variability, causing the detection of a number
of weak sources that rise above the detector sensitivity during flares in
long-exposure satellite experiments, and the many steep-spectrum radio sources
that flatten at high frequencies, as well as the contribution of radio-galaxies,
we find that the level of residual fluctuation due to discrete extragalactic
foreground sources could be factor of $\sim 2 - 3$ higher than the above
estimate.
We show that the Blazar induced fluctuations contaminate the CMB spectrum at the
level of $\approx 20-50 \%$ at $\ell = 500$ and $50-100\%$ at $\ell = 800$, and
thus they cannot be neglected in the derivation of the primordial CMB
fluctuation spectrum. Careful cleaning for Blazar contamination of high
sensitivity/high resolution CMB maps is therefore necessary before firm
conclusions about weak features, like secondary high-$\ell $ peaks of the CMB
power spectrum or very weak signals like CMB polarization measurements, can be
achieved.


}

 \maketitle

\section{Introduction}

The Cosmic Microwave Background (hereafter CMB) contains - {\it in nuce} - a
large amount of cosmological information. The paradigm follows from the fact
that the CMB power spectrum is sensitive to the cosmological parameters which
describe the fundamental properties of the universe (see, e.g., Jungman et al.
1996, Bond et al. 1997). The ability of the CMB data to constrain the
cosmological parameters hinges upon the existence of acoustic oscillations in
the primordial plasma at the last scattering epoch and on the ability of CMB
experiments to disentangle such primordial fluctuations from those induced by
cosmic structures at lower redshift.

CMB temperature fluctuations have been measured and studied in detail for the
past decade starting with the large angular scale ($\theta \simgt 9$ deg
corresponding to a scale $\ell \simlt 20$) COBE detection (Smooth et al. 1992)
up to the more recent and small-scale detection by BOOMERANG (deBernardis et al.
2000), MAXIMA (Hanany et al. 2000), DASI (Halverson et al. 2002), VSA (Grainge
et al. 2002), CBI (Pearson et al. 2002) and, lastly, by WMAP (Bennet et al.
2003) which probes the CMB angular spectrum till $\ell \sim 800$ over the whole
sky with unprecedented precision. Extracting constraints on the cosmological
parameters from the measurements of the CMB power spectrum is conceptually
simple, but much more delicate in practice.
In fact, the ability of the present and upcoming CMB experiments to determine
the cosmological parameters requires a careful cleaning of the CMB maps from the
galactic and extra-galactic foregrounds (see, e.g., Refregier \& Spergel 2000).

At high Galactic latitudes point-like sources are one of the main contaminants
at small angular scales where, even when the brightest sources are identified and then
subtracted, their residual contribution may dominate over the cosmic variance
uncertainty.
It has been shown, in fact, that point-like sources are one of the main
foreground issue for the CBI experiment which is sensitive up to $\ell \sim
3500$ (Myers et al. 2002). However, such an issue also applies to other CMB
experiments like VSA, DASI, or WMAP which probe the CMB angular spectrum at
lower $\ell \simlt 1000$ and it is nonetheless crucial for the up-coming space
missions like PLANCK.
Along this line of reasoning, Pierpaoli (2003) has recently shown that
point-like sources induce a contamination of $\sim 20 \%$ level in the CMB power
spectrum reconstruction even at $\ell \approx 200$ in the simulated WMAP map at
$90$ GHz. This level of contamination grows for increasing values of $\ell$
(i.e., for smaller angular scales) reaching a level of $\sim 80 \%$ at $\ell
\approx 300$.
This implies that point-like sources must be successfully identified and
efficiently subtracted off the CMB maps before recovering the true primordial
fluctuation spectrum. Pierpaoli (2003) also showed that bright sources have the
greatest impact on the power spectrum at moderate $\ell$ values ($\ell \sim
400$) while the fainter sources have a greater impact at high-$\ell$ values.

The recent WMAP experiment (Bennet et al. 2003) resolved 208 bright pointlike
sources, the vast majority of which are Blazars. We show in this paper that
recent Blazar surveys, combined with detailed multi-frequency Spectral Energy
Distribution analysis, provide evidence that a much larger number of Blazars and
active galaxies must be present as positive temperature fluctuations already
visible in the available CMB maps with even moderate spectral and spatial
sensitivity. To this end, we discuss in Sect.2 the astrophysical aspect of
Blazars at radio and sub-mm frequencies and we discuss in Sect.3 their
contribution to the BOOMERANG 90/150 GHz anisotropy map.  We also compare our
results with the catalog of bright sources derived from the WMAP team. Based on
an updated derivation of the Blazar counts, we derive in Sect.4 the average sky
brightness of the mm sky due to the Blazar population and we compute in Sect.5
their angular power spectrum compared to the CMB power spectrum. We summarize
our conclusions and we discuss the relevance of our results for the present and
for the next coming CMB precision experiment in the final Sect.6.

\section{Blazars at cm and mm wavelengths}

Blazars are a type of radio loud Active Galactic Nuclei (AGN) distinguished
by their extreme, in some cases unique, properties. These include large amplitude
rapid variability, apparent superluminal motion, high level of polarization and
emission of radiation across the entire electromagnetic (e.m.) spectrum,
from low frequency radio waves to gamma rays, up to TeV energies in some cases.
This peculiar behaviour is thought to be due to physical phenomena observed under
very special circumstances such as the emission of radiation from a relativistic
jet of material ejected by a radio galaxy in the direction of the observer and seen
at small angle with   respect to the line of sight (\cite{urrpad95}, \cite{blarees}).

The broad band e.m. spectrum of Blazars [represented in a $\nu f(\nu)~vs.~\nu$
plane and usually referred to as the Spectral Energy Distribution or SED] is
marked by a flat radio spectrum, widely believed to be due to synchrotron
radiation, whose luminosity increases with frequency until it reaches a maximum
that is normally located at $10^{12}-10^{14}$ Hz but can also reach $\sim
10^{17}$ Hz in some extreme cases. Above the synchrotron peak the spectrum falls
sharply until Inverse Compton (IC) emission emerges and rises again to form a
second luminosity peak which is typically located at frequencies $\nu \approx
10^{17}-10^{19}$ Hz [see, e.g., Fig.\ref{fig.sed} for two examples of Blazar
SEDs; many other examples of Blazar SEDs can be found in the catalog of
\cite{gio02a}].

Because the special conditions that make a radio galaxy appear like a Blazar are
rather improbable, Blazars are much rarer than other AGN. However, since
Blazars emit over the entire e.m. spectrum, while other types of AGN emit most
of their radiation in the radio (radio loud only), infra-red, optical, and X-ray
band, Blazars remain the extragalactic sources more frequently found in the
remaining parts of the e.m. spectrum, that is the Gamma-Ray, the TeV and the
microwave bands.

There are two types of Blazars: those showing a featureless spectrum,
called BL Lacs, which are rare
and do not show a strong cosmological evolution, and the flat radio spectrum
QSOs (FSRQ), which have broad lines in their optical spectrum, are much more
abundant and evolve strongly on cosmological times (\cite{urrpad95},
\cite{pad02}).
Intensity and spectral variations are conspicuous at optical frequencies and
become even larger at higher frequencies. Variability in the radio part of the
e.m. spectrum, as demonstrated by a monitoring program lasting more than 25
years, can reach factors of $\sim$ 2 to 10 on typical time-scales of weeks to
years with a clear tendency to become more pronounced and more frequent at
higher frequencies; polarization at radio frequencies is often present at a
level of a few percent and in some case up to 10\% (see, e.g., \cite{hughes92}
and the on-line UMRAO database
http://www.astro.lsa.umich.edu/obs/radiotel/umrao.html).

Radio sources of this type display a spectral shape
which is not too different from that of the CMB spectrum before and around the
peak of its power in the frequency range $\sim 50-200$ GHz (see, e.g., Fig.1).
For this reason Blazars can easily contaminate CMB maps (e.g., they can be
confused with CMB fluctuations) if they are bright enough to be detected by CMB
experiments. The recent results of the WMAP satellite experiment (\cite{ben03a})
clearly demonstrate that Blazars are present in large numbers in CMB fluctuation
maps. In fact, we found, from a cross-correlation of the 208 sources of the WMAP
catalog with catalogs of known sources and with the NED database, that $\sim 85
\%$ of the WMAP sources are Blazar-like objects. Specifically, we found 139
FSRQ, 24 BL Lacs, and 15 radio galaxies. In addition to this evidence, we found
that the three extragalactic sources detected well above the sensitivity
threshold of the BOOMERANG maps (\cite{deber00}, \cite{cob03}) are also
Blazars.\\
To produce a more quantitative analysis of the contamination of CMB anisotropy
maps by Blazars, we provide in the following, a list of the bright known Blazars
which can be found in the 90/150 GHz BOOMERANG map. We choose to analyze the
BOOMERANG maps because it covers a statistically representative large area of the sky $\sim
1800~deg^2$) and because they are available (even though in a low
resolution format) on the web (see, e.g., http://cmb.phys.cwru.edu/boomerang/).

\section{Blazars in the BOOMERANG CMB anisotropy maps}

We search here for Blazars that might have already been detected in the
BOOMERANG maps but not yet recognized as such. To this aim, we have built the
SED of all known Blazars for which that are enough published measurements (taken
from NED and from various other surveys) to allow for a reasonably safe
extrapolation of the flux intensity at the frequencies 90 and 150 GHz of the
BOOMERANG maps.
\begin{figure}
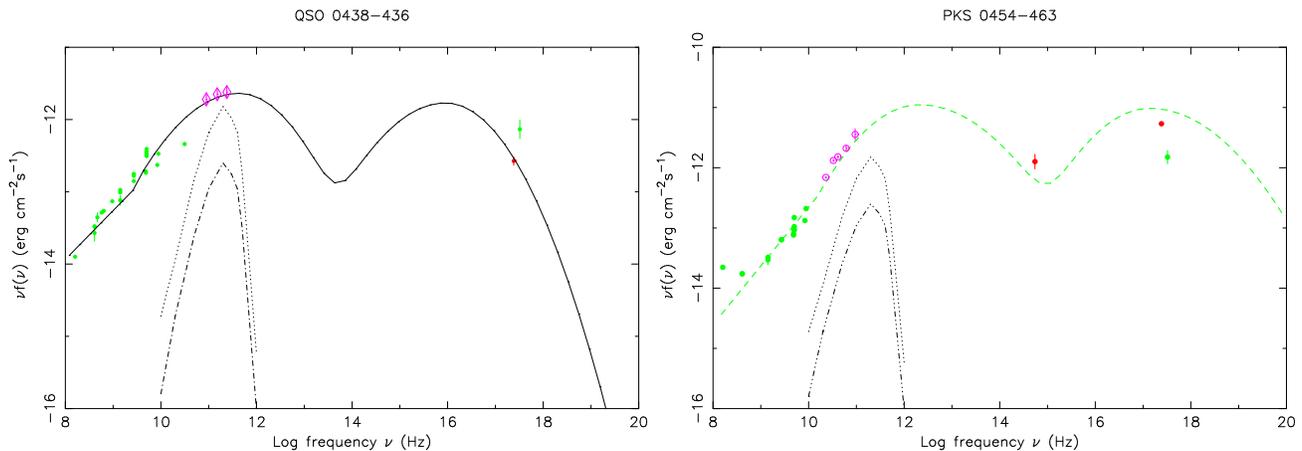

\centerline{
 \hbox{
 \psfig{figure=sed_J0440m4333.ps,width=8.5cm,angle=-90}
 \psfig{figure=sed_J0455m4615.ps,width=8.5cm,angle=-90}
 }}
\caption{{\bf Left} We show the SED  of the FSRQ source,PKS 0438-436, one of the
three Blazars reported by the original authors to have been detected in
BOOMERANG maps (deBernardis et al. 2000, \cite{cob03}). The three open diamonds
at 90, 150 and 240 GHz are, in fact, BOOMERANG measurements taken from
\cite{cob03}. The dotted and dashed-dotted curves below the data points,
represent the frequency spectrum of CMB fluctuations of the order of 300 and 50
$\mu K$, respectively. The solid curve shows the SED in the $\nu - f_{\nu}$
plane as expected from the Synchrotron Self Compton (SSC) model.
{\bf Right} We show here the SED of the FSRQ source PKS 0454-463, whose expected
flux at mm frequencies, estimated from the SSC model and shown as a dashed
curve, is well above the BOOMERANG detection limit. This extrapolation is
confirmed by the WMAP detection of this source at 23, 33, 41, 61 and 94 GHz
(\cite{ben03b}) which are shown as open points. The dotted and dashed-dotted
curves below the data points, represent again the frequency spectrum of CMB
fluctuations of the order of 300 and 50 $\mu K$, respectively. }
 \label{fig.sed}
\end{figure}
We started our analysis from the few point-like sources that have already been
recognized in the BOOMERANG maps (\cite{deber00}). Their follow-up (Coble et al.
2003) has shown their galactic (6 out of 9 sources) or extra-galactic (3 out of
9 sources) nature. The three extragalactic foreground sources are namely a
Blazar (PKS0537-441), a BL Lac (PKS0521-365) and a Flat Spectrum Radio QSO
(PKS0438-43). For one of these, namely PKS0438-436, we report the SED in
Fig.\ref{fig.sed} where filled circles represent NED measurements and the three
open diamonds at 90, 150 and 240 GHz represent the BOOMERANG measurements taken
from \cite{cob03}. To obtain a flux estimate around the peak of the CMB we
fitted the multi-frequency points with a Synchrotron Self Compton (SSC) model
(\cite{perri02}, see also Perri 2003, PhD Thesis, University of Rome, La
Sapienza) which is a very good approximation of the SED of Blazars in the radio
and mm band. The dotted and dashed-dotted curves in Fig.\ref{fig.sed} represent
the spectral distribution of typical CMB temperature fluctuations of the order
of 300 and 50 $\mu K$ respectively, which can be clearly detectable in the
BOOMERANG maps. Based on the SED shown in Fig.\ref{fig.sed}, the source
PKS0438-436 is then associated with a fluctuation of order $\simgt 300 ~\mu K$,
as already recognized in the BOOMERANG maps. The SED of the other sources
PKS~0521-365 and PKS~0537-441 are similar to the previous case but are not
reported here for the sake of brevity.

Following a similar procedure we found more than 50 Blazars, or Blazar
candidates, selected through a multi-frequency technique like that used in the
DXRBS (\cite{lan01,pad02}) using the NVSS (radio), GSC2 (optical)  and the Rosat
(X-ray) all sky survey, for which the extrapolation of the radio data is above
the detection limit of the BOOMERANG experiment.
Since the SED of Blazars allows for fluxes above the BOOMERANG sensitivity and
below the Rosat all sky survey sensitivity limit, the number of objects found is
only a lower limit.

As an example of our findings, we show in the right panel of Fig.\ref{fig.sed} the
SED of the source PKS~0454-483 whose expected emission at mm frequencies is
similar to that of the three Blazars already found in the BOOMERANG map.
Whenever a Blazar was found to have a flux corresponding to a CMB fluctuation
significantly above $50 ~\mu$K, we have identified its position in the 150 GHz
BOOMERANG map.

\begin{figure}
\centerline{
 \vbox{
 \psfig{figure=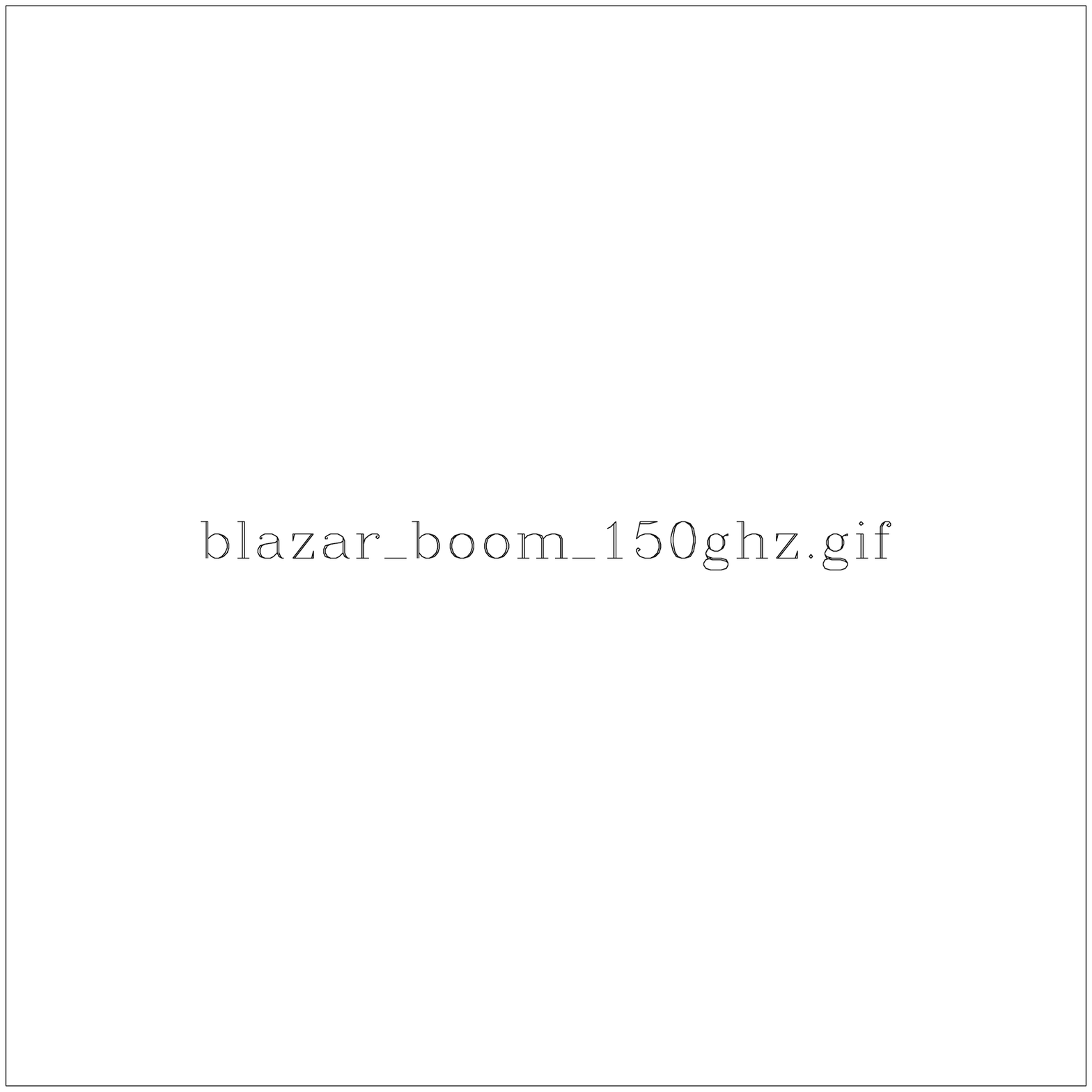,width=18.0cm,angle=0}
 }}
\caption{The 150 GHz CMB fluctuation map (taken from
http://cmb.phys.cwru.edu/boomerang/) obtained from BOOMERANG is here correlated
with the positions of known and candidate high galactic latitude ($|b| \ga 20$)
Blazars (and radio galaxies) whose expected flux at 150 GHz (estimated from
lower frequency archive data measurements) is high enough to be above the
BOOMERANG detection limit, and should appear as green/yellow or red areas (see
Table 1. See also http://www.asdc.asi.it/boomerang/ for the full list of FSRQ
SEDs). Some of these objects (12 out of the 54 Blazars we list in Tab.1) are
also included in the WMAP source catalog.
}
 \label{fig.BOOMERANGmap}
\end{figure}

We report in Fig.\ref{fig.BOOMERANGmap} the location of the whole set of Blazars
found in the 150 GHz BOOMERANG map. Their sky coordinates (RA and DEC), radio
flux, optical V-band magnitude and redshift are also reported for completeness
in Tab.\ref{tab.BOOMERANGsources}.
\begin{table}[h]
\caption{The list of flat-spectrum radio sources (mostly Blazars) found in the
BOOMERANG maps. For each object we list the source name (col.1), the coordinates
Ra (col.2) and Dec (col.3), the radio flux at 5 GHz in units of mJy (col.4), the
optical V-band magnitude (col.5), the redshift taken from NED (col.6). Some
notes on the listed sources are added in col.7 of the table. }
\begin{center}
\begin{tabular}{lrrrccl}
\hline \multicolumn{1}{c}{Source name} & \multicolumn{1}{r}{Ra(J2000.0)} &
\multicolumn{1}{r}{Dec(J2000.0)} & \multicolumn{1}{c}{Radio Flux} &
 \multicolumn{1}{c}{V-mag} &
 \multicolumn{1}{c}{Redshift} &
 \multicolumn{1}{c}{Notes} \\
\hline
 PKS 0252-549    &02 53 30.7&-54 41 40.9&1193&17.7&0.537&FSRQ - WMAP\\
 PKS 0257-510&02 58 37.7&-50 52 15.9&452&23&0.834&FSRQ \\
 PKS 0308-611 &03 09 55.6&-60 58 27.4&1103&18.6&-& NED: QSO \\
 PKS 0310-558&03 12 05.8&-55 41 38.0&501&18&-& Blazar candidate\\
 PKS 0317-570&03 18 55.9&-56 50 52.0&257&17.5&-& Blazar Candidate\\
 PMN J0321-3711  &03 21 23.2&-37 11 33.0&5020&-&-&NED: Radio S. Extended - WMAP\\
 PKS 0340-372 &03 42 05.4&-37 03 10.9&872&18.1&0.284& QSO \\
 PKS 0402-362&04 03 53.0&-36 04 47.9&1132&17.2&1.417& FSRQ \\
 PKS 0405-385    &04 06 58.7&-38 26 22.9&830&17.7&1.285&FSRQ - WMAP\\
 PKS 0410-519 &04 11 36.1&-51 49 19.9&361&17.5&-& NED: QSO \\
 PMN J0419-3010&04 19 48.2&-30 10 06.9&184&17.5&-& Blazar candidate\\
 PMN J0422-3844&04 22 14.6&-38 44 48.1&130&17&3.11& NED: QSO \\
 WGA J0424.6-3849&04 24 39.2&-38 49 01.9&309&18.5&2.34&FSRQ \\
 PKS 0422-380    &04 24 41.8&-37 56 23.9&1706&18.1&0.782&FSRQ - WMAP \\
 WGA J0428.8-3805&04 28 50.4&-38 05 44.1&51&16.5&0.15&BL Lac \\
 1RXS J043208.7-35065&04 32 08.6&-35 06 51.1&182&18&-& Blazar candidate\\
 PKS 0432-606&04 33 34.0&-60 30 14.0&636&19&-& NED: QSO \\
 PKS 0435-300&04 37 37.4&-29 54 11.1&691&17.2&1.328& FSRQ \\
 0438-43 &04 40 17.7&-43 32 56.0&3933&18.8&2.852&FSRQ - WMAP \\
 PKS 0439-331&04 41 34.6&-32 59 52.0&197&-&-& NED: Blazar candidate \\
 1RXS J044510.0-38382&04 45 10.0&-38 38 24.0&244&17&-& NED: QSO - PKS 0443-387\\
 PKS 0448-392 &04 49 42.0&-39 10 51.9&819&16.5&1.288& FSRQ \\
 PKS0450-469&04 51 52.5&-46 53 02.0&541&18.3&-& NED: QSO \\
 1Jy0454-463     &04 55 51.2&-46 16 01.9&1653&17.4&0.858&FSRQ - WMAP\\
 PKS 0506-61     &05 06 43.6&-61 09 39.9&1211&16.9&1.093&FSRQ - WMAP\\
 NGC 1808&05 07 43.0&-37 30 29.8&238&12.6&0.003& Starburst gal.\\
 PMN J0510-3533&05 10 54.0&-35 33 09.0&123&18.1&-& Blazar candidate\\
 PKS 0511-484 &05 12 51.4&-48 24 09.0&1094&19.8&0.306& NED: Radio galaxy \\
 Fairall 790&05 14 35.4&-49 03 33.1&310&15.9&0.09& NED:  Elliptical gal.\\
 1Jy 0514-459    &05 15 48.1&-45 57 00.0&990&17.5&0.194&FSRQ \\
 PKS 0518-45         &05 19 49.0&-45 46 45.8&14999&15.8&0.035&FSRQ - WMAP \\
 PKS 0521-365    &05 22 59.4&-36 28 01.9&8180&14.6&0.055&BL Lac - WMAP \\
 PMN J0525-3343&05 25 04.9&-33 43 00.1&210&18.7&4.4& FSRQ \\
 PMN J0529-3555 &05 29 37.1&-35 55 22.0&329&17.7&-& Blazar Candidate\\
 1RXS J053132.1-35333&05 31 32.0&-35 33 30.9&173&18.5&0.9& Blazar candidate\\
 PKS 0532-378&05 34 18.0&-37 47 07.0&528&19.5&1.668& NED: QSO \\
 PKS 0534-340&05 36 28.5&-34 01 12.0&663&17.8&0.683& FSRQ \\
 1Jy0537-441     &05 38 51.0&-44 05 06.0&4805&15&0.896&BL Lac - WMAP \\
 PKS 0539-543    &05 40 47.2&-54 18 20.8&373&18.2&-&Blazar Candidate - WMAP \\
 1RXS J054329.1-39563&05 43 29.1&-39 56 38.0&174.8&17.9&-& Blazar candidate\\
 PMNJ0545-4757 &05 45 06.4&-47 57 11.9&294&17&-& Blazar Candidate in Cluster\\
 PKS 0549-575    &05 50 08.1&-57 32 30.8&355&19.5&-&QSO from NED - WMAP \\
 PKS 0548-322&05 50 40.7&-32 16 18.8&213&15.5&0.069&BL Lac     \\
 PKS 0548-317&05 50 47.4&-31 43 59.8&1034.5&9.9&0.033&BL Lac \\
 PKS 0555-374 &05 57 11.1&-37 28 13.0&152&14.2&0.0044& Elliptical in Cluster\\
 1RXS J060414.4-31560&06 04 14.4&-31 56 02.0&2960.4&18.5&0.452&FSRQ \\
 PMN J0606-3448 &06 06 08.1&-34 47 43.0&135&17.9&2.28& NED: QSO\\
 PKS 0607-605&06 07 55.0&-60 31 55.9&558&18.2&-& NED: QSO \\
 PKS 0606-306&06 08 40.9&-30 41 38.0&100&16.9&-& Blazar Candidate \\
 PKS 0610-316&06 12 29.2&-31 39 14.0&643&18.5&-& NED: QSO \\
 PKS 0613-312 &06 15 16.8&-31 16 59.8&173.6&17&-& NED: QSO \\
 PKS 0410-519 &06 22 02.1&-47 35 52.0&315&13&-& Blazar Candidate \\
 WGA J0631.9-5404 &06 31 59.5&-54 04 31.0&155&18.2&0.193&FSRQ \\
 PKS 0646-437 &06 48 16.7&-43 47 00.9&126&18&1.029&FSRQ \\
 \hline
\end{tabular}
\label{tab.BOOMERANGsources}
\end{center}
\end{table}
To summarize, we found 54 known Blazars with flux greater than $\sim $ 200 mJy
that are probably associated with fluctuations of amplitude $\simgt 50 \mu K$,
in the 150 GHz BOOMERANG map. The number of these Blazar point-like sources is a
factor $> 15$ larger than previously reported (\cite{cob03}). A much larger
number of fainter Blazars are evidently expected to be associated with lower
amplitude ($\simlt 50 \mu K$) CMB fluctuations due to the fact that the Blazar
counts steeply increase at flux fainter than $200$ mJy, as will be discussed in
the next Sect.4.

The analysis of the BOOMERANG map is just an example of what we could find in a
specific area of the sky surveyed at mm wavelenghts. In particular, we estimate
that more 200 Blazars should be found above the BOOMERANG sensitivity limit in
1000 $deg^2$ area. The WMAP anisotropy maps have basically the same sensitivity
threshold and spatial resolution of the BOOMERANG maps and thus our results can
be quite easily rescaled to the WMAP all-sky survey. Given the previous results,
the Blazar populations are expected to produce a non-negligible contamination of
the CMB maps as well as of the CMB power spectrum. We will address this point
more specifically in the next Sects.4 and 5.

\section{The sky brightness due to Blazars at cm and mm wavelengths}

Precise estimates of the contribution of Blazars to the sky brightness at radio
and mm frequencies have so far been hampered by the lack of sufficiently deep
surveys. For a long time only very shallow (1 or 2 Jy) surveys have been
available at cm frequencies (\cite{sti91}, \cite{wp85}). However, the recent
multi-frequency approach to Blazar search is changing the picture (e.g.
\cite{mue98}, \cite{gio99}, \cite{pad02}, \cite{gio02b} \cite{gio02c}), making
it possible to detect many fainter Blazars.

The integrated radio flux due to a population of discrete sources in the flux
range $S_{min}-S_{max}$ can be calculated as follows:
 \be
 \displaystyle {I=\int_{S_{min}}^{S_{max}} S~{dN \over dS}~dS } ~,
 \label{eq.lognlogs}
 \ee
where the quantity $dN/dS$ is usually referred to as the differential LogN-LogS
distribution that, for a population of sources with constant density, is given
by a power-law with slope equal to 2.5 which then flattens at faint fluxes of
order of $\simlt 1-10$ mJy, depending on the Blazar luminosity function. The
LogN-LogS of BL Lacs is presently known down to $\approx$ 3.5 mJy for the
subsample of extreme High Energy Synchrotron peaked (HBL) sources (\cite{gio99},
\cite{gio02b}) while the best estimate of the LogN-LogS of the full population
of BL Lacs and FSRQs currently comes from the Deep X-Ray Blazar Survey (DXRBS,
\cite{lan01}). This is the deepest complete and radio-flux limited Blazar survey
presently available which extends to 50 mJy (\cite{pad03}).
At radio frequencies, FSRQs are more abundant than BL Lacs by about a factor of
$\approx 4$ and show a strong cosmological evolution, while BL Lacs either do
not evolve or show negative cosmological evolution (\cite{pad02}, \cite{gio99},
\cite{bad98}, \cite{rec00} ). The LogN-LogS of BL Lacs is approximately
euclidean up to at least 50 mJy, while below 20 mJy it flattens substantially,
at least in the subsample of extreme HBL BL Lacs (\cite{gio02b}). From the DXRBS
survey we know that the surface density of sources brighter than 0.1 Jy at 5~GHz
is 0.1 object per square degree for FSRQs and 0.025 objects per square degree
for BL Lacs (\cite{gio02b}, \cite{pad03}). We can then assume a Blazar overall
density of 0.125 $deg^{-2}$ with flux $S \ge 0.1 $ Jy.
\begin{figure}
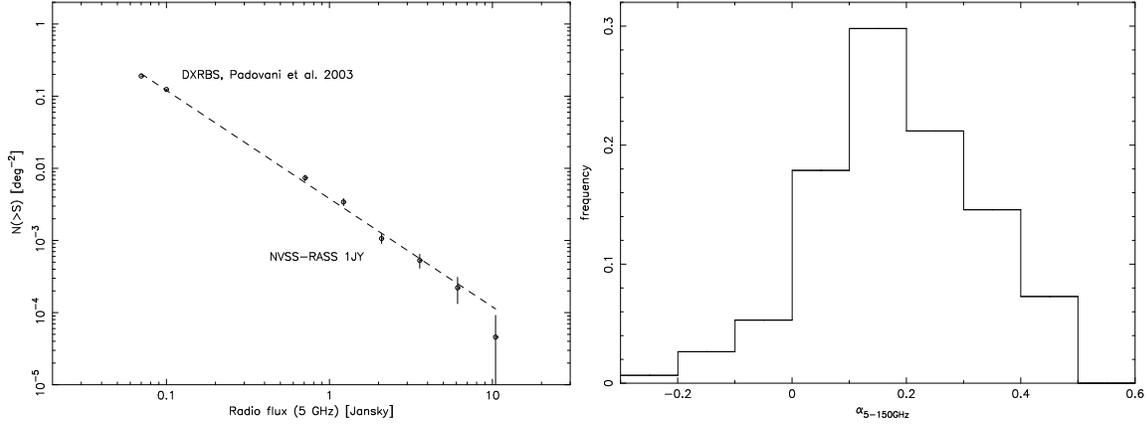

\centerline{
 \hbox{
 \psfig{figure=logns_allblazars_5ghz_new.ps,width=7.5cm,angle=-90}
 \psfig{figure=blazar_slopes.ps,width=7.5cm,angle=-90}
 }}
 \caption{{\bf Left}. The 5 GHz (integral) LogN-Logs of Blazars
 built with data from the NVSS-RASSS 1Jy survey and the DXRBS point at 0.1 and 0.05 Jy.
 The dashed line is a power-law with slope 1.6.
 {\bf Right}. We show  here the distribution of spectral slopes between cm and mm
 wavelenghts from the sample of 135 Blazars of the NVSS-RASS 1Jy survey with
 $\alpha < 0.5 $.}
 \label{fig.counts}
\end{figure}
Since FSRQs (the dominant Blazar population) evolve strongly, while BL Lacs show
no evolution (or slightly positive) at high radio fluxes (\cite{pad02}), and
negative evolution below 20 mJy (\cite{gio99}, \cite{gio02b}), we can
conservatively assume an overall Blazar differential LogN-LogS with an
"Euclidean" slope of 2.5  which extends up to a break flux of the order of 10
mJy below which it flattens significantly to a slope of $\sim 2.2$. This
assumption is fully supported by the LogN-logS (see Fig.\ref{fig.counts}) that
we have built using a sample of 135 Blazars (BL Lacs + FSRQs) from the NVSS-RASS
1Jy and the 0.1 and 0.05 Jy points of the DXRBS (\cite{pad03}). While this LogN-logS is
consistent with the counts of flat spectrum radio sources derived by
\cite{tof98} at $\approx 1$ Jy, it does not show any flattening until a rather
faint radio flux is reached and not just below 1 Jy as in the predictions based
on \cite{tof98}.

Thus, given the  previously discussed LogN-LogS, and setting consistently the
lower integration limit to 0.1 mJy,  Eq.(\ref{eq.lognlogs}) yields a radio
brightness of $ \approx 0.7-1.0~Jy/deg^2$ at 5 GHz. In order to properly
extrapolate this brightness to higher frequencies, we have to make use of the
distribution of radio to mm spectral slopes for the Blazar population. To
estimate realistic slope values we have built the Spectral Energy Distribution
(SED) of all Blazars in the NVSS-RASS 1Jy survey (\cite{gio02c}) using the many
radio measurements available from the literature (mostly taken from NED). Since
all objects in the NVSS-RASS survey are brighter than 1 Jy, several radio data
points are normally available for each Blazar over a wide frequency range. We
have then fitted a a power law to the radio part of spectrum and measured the
distribution of spectral slopes $\alpha $ [here $f(\nu)\propto \nu^{-\alpha}$]
between the cm (e.g., $\approx 5 $GHz) and  mm region (e.g., $\approx 41, 90,
94$ and $150$ GHz). We have derived the normalized distribution shown in
Fig.\ref{fig.counts} using the subsample of 135 objects with $\alpha < 0.5 $
which is a requirement normally applied to build Blazar samples. Using such a
spectral slope distribution, the expected 150 GHz sky brightness due to Blazars
ranges from $\approx 7\times 10^{10}$ to $\approx 1\times 10^{11}
Jy~Hz~deg^{-2}$ depending mildly on the assumptions for the LogN-LogS
parameters. Since the CMB brightness at 150 GHz is $\approx 2.1~10^{16}
Jy~Hz~deg^{-2}$ (e.g. \cite{has00}), the Blazars brightness is  $\approx
(3-5)\times 10^{-6}$ that of the CMB, which is only a factor of a few below the
level of the presently measured CMB fluctuations and similar to the sensitivity
planned for the upcoming PLANCK mission. At 41 and 94 GHz we obtain a sky
brightness of 3-5 $\mu K$ assuming that the slope of the LogN-LogS flattens from
2.5 to 2.1 at 10 mJy and integrating it down to 1 and 0.1 mJy, respectively.

This estimate is only a lower limit to the average sky brightness induced by
Blazars since the DXRBS and NVSS-RASS 1Jy surveys only include sources with flat
radio spectrum between 1.4 and 5 GHz, while it is well known that a number of
radio source spectra that are steep at low frequencies progressively flatten at
higher frequencies. Through direct 15 GHz source counts measurements (see Taylor
et al. 2003) estimate that the contribution of steep-spectrum sources at 1.4-5
GHz increase the extrapolated source counts at high frequency by a factor of
$\sim 2$.

In addition, other sources that are not accounted for in Blazar surveys have
been detected, as a minority population but still in fair numbers, in the WMAP
experiment. These are, e.g., radio galaxies which have a often extended emission
with steep radio spectrum, that outshines a much flatter spectrum  nuclear
component which may emerge and dominate the emission above a few tens of GHz.
One example of such sources is Pictor A, a bright radio galaxy with extended
steep-spectrum  emission and a flat-spectrum  Blazar like, nuclear component
(Fiocchi \& Grandi 2003, in preparation), which is included in the WMAP source
catalog and is also present in the BOOMERANG maps (see Sect.2).

\section{The angular power-spectrum contributed by Blazars}

Assuming that the Blazar are spatially distributed like a Poissonian random
sample, then the angular power spectrum, $C_{\ell,Blazar}$, contributed by these
sources is given by
 \be
 C_{\ell,Blazar} = \int_{S_{min}}^{S_{max}} dS~{dN \over dS}~S^2 ~,
 \label{Eq.cl}
 \ee
(Tegmark and Efstathiou 1996, Scott and White 1999). The quantity at right hand
side in Eq.(\ref{Eq.cl}) is the usual Poisson shot-noise term (Peebles 1980,
Tegmark and Efstathiou 1996) and a further term, $\omega(I)^2$, adds to it if
the clustering of sources is not negligible (Scott and White 1999). This last
clustering term can be even a factor $\sim 10$ larger than the Poissonian term
at large and intermediate angular scales, $\ell < 1000$ (Scott and White 1999).

\begin{figure}
\centerline{
 \hbox{
 \psfig{figure=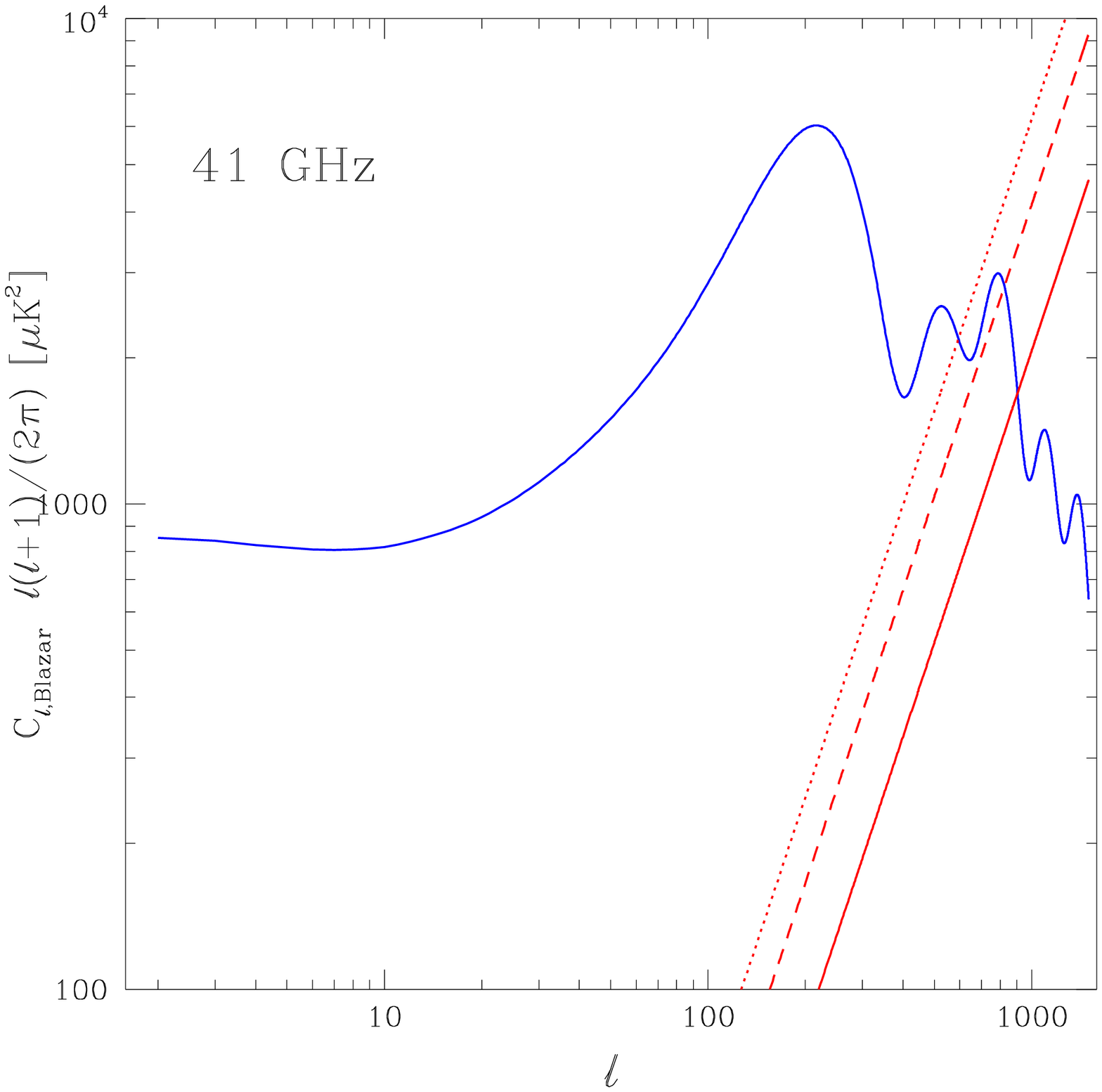,width=8.0cm,angle=0}
 \psfig{figure=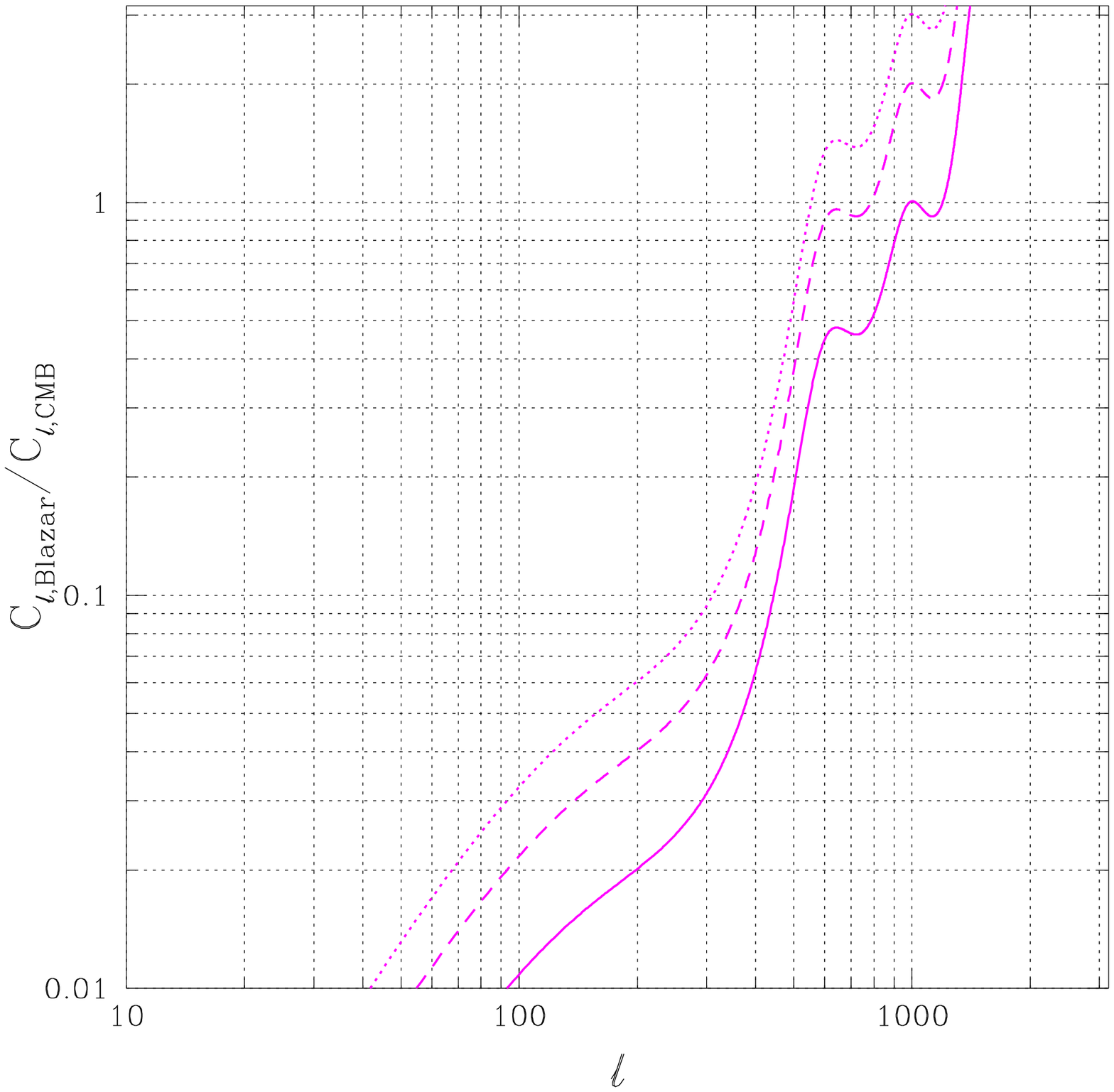,width=8.0cm,angle=0}
 }
 }
 \caption{{\bf Left}. We show the contribution of Blazars to the CMB fluctuation spectrum in the
WMAP Q channel at 41 GHz as evaluated from the LogN-LogS given in Sect.4 (solid
line). We also show the angular power spectrum for the Blazar population by
adding the contribution of radio sources with steep-spectrum at low radio
frequencies which flatten at higher frequencies (dashed line). The dotted line
includes also the effect of spectral and flux variability (see text for
details). A typical CMB power spectrum evaluated in a $\Lambda$CDM cosmology
with $\Omega_m=0.3, \Omega_{\Lambda}=0.65, \Omega_b=0.05$ which best fits the
available data is shown for comparison. {\bf Right}. We show here the ratio
$C_{\ell,Blazar}/C_{\ell,CMB}$ between the Blazar power spectra and the CMB
power spectrum.
 }
 \label{fig.cmb.spec.41}
\end{figure}

For the Blazar population described by the LogN-LogS given in
Fig.\ref{fig.counts} we found $C_{\ell,Blazar} \approx 31 Jy^2 sr^{-1}$ at 41
GHz and $C_{\ell,Blazar} \approx 30 Jy^2 sr^{-1}$ at 94 GHz. Our results can be
translated into temperature units, as is usual in CMB anisotropy studies, using
the conversion between the isotropic black-body (Planckian) brightness
$B_0(\nu)$ and the CMB temperature $T_0$ which writes as
\be
 {\partial B_0 \over \partial T_0} = {k \over 2} \bigg( {k T_0 \over h c}
 \bigg)^2 \times \bigg[ {x^2 \over sinh(x/2)}\bigg]^2
 = \bigg( {24.8 Jy ~sr^{-1} \over \mu K} \bigg)
 \times \bigg[ {x^2 \over sinh(x/2)}\bigg]^2 ~,
 \label{Eq.conv}
 \ee
 where $x \equiv h \nu/k T_0 = \nu / 56.84 GHz$ is the a-dimensional frequency given in
terms of the Planck constant $h$, of the speed of light $c$ and of the Boltzmann
constant $k$.
For WMAP, we found $C_{\ell,Blazar} \approx 1.3\cdot 10^{-2} ~\mu K^2 sr$ at 41
GHz and $C_{\ell,Blazar} \approx 6.5 \cdot 10^{-4} ~\mu K^2 sr$ at 94 GHz. In
Fig.\ref{fig.cmb.spec.41} we plot the quantity $C_{\ell,Blazar} \ell
(\ell+1)/(2\pi)$ and we compare it to the CMB fluctuation power spectrum which
best fits the available data.
The solid line in Fig.\ref{fig.cmb.spec.41} shows the contribution of Blazars to
the CMB fluctuation spectrum in the WMAP Q channel at 41 GHz as evaluated from
the LogN-LogS given in Eq.\ref{eq.lognlogs} which yields $C_{\ell,Blazar}
\approx 1.3\cdot 10^{-2} ~\mu K^2 sr$. We stress here that the previous value is
a definite lower limit for $C_{\ell,Blazar}$ since it neglects the contribution
of steep-spectrum sources at low frequencies which flatten at these frequencies
(41 and 94 GHz) and the effect of flux variability. For this reason we also show
a more realistic case for the Blazar power spectrum by adding the contribution
of radio sources with steep-spectrum at low radio frequencies which flatten at
higher frequencies (the dashed line in Fig.\ref{fig.cmb.spec.41}). These sources
likely provide a value of $C_{\ell,Blazar}$ which is higher by a factor $\sim 2$
(see also discussion in Taylor et al. 2003). The effect of spectral and flux
variability also allows for an increase of the fluctuation level
$C_{\ell,Blazar}$ because many Blazars which are below the sensitivity threshold
of CMB experiments can be detected during flares. The Blazar flux variability at
millimeter wavelengths may be very substantial (higher than the factors 3-10 on
time scales of weeks to years seen at cm wavelengths) and could definitely
increase the level of contamination of CMB maps when these are built over long
integration periods. The dotted line in Fig.\ref{fig.cmb.spec.41} shows the
Blazar power spectrum when also this last contribution is taken into account.

To better quantify the level of contamination of the CMB power spectrum induced
by Blazars we show in the right panel of Fig.\ref{fig.cmb.spec.41} the ratio
$C_{\ell,Blazar} /C_{\ell,CMB}$  where $C_{\ell,CMB}$ has been evaluated in a
$\Lambda$CDM cosmology with $\Omega_m=0.3, \Omega_{\Lambda}=0.65, \Omega_b=0.05$
which best fits the available data. It is evident from our results that, even in
the case of the lower limit to $C_{\ell,Blazar}$ (solid curve in
Fig.\ref{fig.cmb.spec.41}), the population of flat spectrum radio sources
produces a power spectrum which amounts to $\sim 20 \%$ of the CMB power
spectrum at $\ell \approx 500$ and reaches $\sim 50 \%$ of the CMB power
spectrum at $\ell \approx 600$. In this case the Blazar power spectrum equals
the CMB power spectrum at $\ell \simlt 10^3$. In a more realistic case, the
addition of steep-spectrum radio sources at low-$\nu$ which flatten at higher
$\nu$'s makes the contamination of the primordial CMB spectrum more significant.
In fact, in this last case, the Blazar power spectrum (dashed curve in
Fig.\ref{fig.cmb.spec.41}) reaches a level $\sim 40 \%$ at $\ell \approx 500$
and nearly reaches equality at $\ell \approx 600$. Inclusion of variability
effects increases even more the contamination (dotted curve in
Fig.\ref{fig.cmb.spec.41}) reaching a level of $\sim 60 \%$ at $\ell \approx
500$ and equalizing the CMB power spectrum at $\ell \approx 550$. Here we have
not considered the statistical uncertainties of the CMB power spectrum data.
These results show that Blazar-like (flat-spectrum) radio sources introduce a
non-negligible contamination in CMB maps and in the CMB power spectrum
reconstruction, even in the lower-limit case of considering just the
flat-spectrum sources at high frequencies. Notice, that the Blazar contamination
of CMB maps decreases with increasing frequency and it becomes $\simlt $ a few
\% at $\ell \simlt 500$ for $\nu \simgt 90$ GHz.

We also stress that the previous calculations are performed neglecting the
clustering term $\omega(I)^2$ and thus they must be considered again as a lower
limit to the realistic angular power spectrum contributed by the Blazars. Such a
clustering term could produce even a factor $\sim 10$ amplification of the power
spectrum at any $\ell < 10^3$ (see, e.g., Scott and White 1999). We will
consider more specifically this additional contribution elsewhere (Giommi \&
Colafrancesco in preparation).

It stems from the previous analysis, that the low-$\ell$ part of the
reconstructed CMB power spectrum is heavily contaminated by Blazars even at
$\ell \sim 400$, where the second and third acoustic peaks are expected in the
best-fit cosmological model (i.e., a scenario with $\Omega_m \approx 0.3,
\Omega_{\Lambda}\approx 0.65, \Omega_b\approx 0.05$). Thus,  the non-negligible
Blazar contamination of the CMB anisotropy spectrum at $\ell \simgt 400$ has
evidently an impact on the derivation of the cosmological information which is
contained in the CMB anisotropy spectrum.
The second acoustic peak at $\ell \approx 500$ is considered as an
incontrovertible evidence of inflationary sound waves (see, e.g., Hu \& White
1997) and its amplitude is mainly sensitive to the values of $\Omega_b h^2$ and
$\Omega_m h^2$. Increasing $\Omega_b h^2$ has the effect of decreasing the
amplitude of the second peak relative to the first and third peaks. A
contamination of $\sim 20 - 50 \%$ at $\ell \approx 500$ would tend to decrease
the power of the CMB spectrum at the position where the second peak is expected
in a $\Lambda$ CDM scenario. This effect, by itself, would bias the quantities
$\Omega_b h^2$ and/or $\Omega_m h^2$ toward high values.
The third acoustic peak (and in principle all the acoustic peaks) is mainly
depending on the amount of Dark Matter in the universe and its amplitude
decreases for increasing values of $\Omega_m h^2$. The high level of
contamination ($\sim 50 - 100 \%$) of the CMB power spectrum at $\ell \approx
800$ as well as the intrinsic uncertainty of the data render the determination
of the third acoustic peak problematic.
The first acoustic peak at $\ell \approx 300$ is contaminated at a level $\simlt
10 \%$ and hence the constraint on $\Omega_0$ set by CMB anisotropy data are
little affected by the Blazar contamination, unless a substantial clustering of
these sources can be considered as realistic. The present data do not provide
strong insights on this last point and we will study it in more details in a
forthcoming paper (Giommi \& Colafrancesco, in preparation).

Finally, we stress that the Blazar emission is significantly polarized at the
level of $\sim 2-10 $ \% at radio and optical frequencies. As a consequence of
the non-negligible Blazar contamination of the CMB maps, the primordial
polarization patterns could be heavily contaminated by these polarized sources.
Hence, a significant contamination of CMB fluctuation maps and the
non-negligible fraction of residual unresolved sky brightness will complicate
the detection of primordial polarization fluctuations, as these are expected to
be present only at a very low level, i.e. at a level $\sim 20 - 50$ lower than
the primary CMB fluctuations (see, e.g., Hu \& White 1997).

\section{Conclusions}

In this paper we studied the contamination of CMB fluctuation maps induced by
flat-spectrum extragalactic radio sources which shine in the frequency bands
probed by BOOMERANG and WMAP. Blazars have in fact a typical spectrum which is
similar to the CMB spectrum in the range 40 - 150 GHz. We predicted the level of
brightness and of the induced CMB anisotropy using the most updated version of
the Blazar LogN-LogS.\\
In view of the results presented in the previous sections, the main conclusion
of this paper can be summarized as follows:

\begin{itemize}
\item Blazars are the major point-like, bright source contaminant of CMB maps.

\item We expect more than 100 Blazars to be present in the BOOMERANG 90/150 GHz
maps above the sensitivity limit. We already identified more than 50 of these
sources on the basis of a detailed analysis of known and candidate Blazars. We
have also identified a large number of Blazars (178 out of a total of 208
sources) among the sources of the WMAP bright source catalog.

\item The importance of Blazars as contaminants of CMB fluctuation maps stems from
the fact that their LogN-LogS is steeper than that of flat spectrum radio
sources previously used to estimate the contamination of CMB maps by
extragalactic sources. The observed Blazar LogN-LogS at 5 GHz combined with the
measured spectral slope distribution implies a LogN-LogS at higher frequencies
which is steeper than Euclidean at least down to 50 mJy at all frequencies
relevant for CMB experiments. We expect, in fact, $\sim 0.4$ sources per deg$^2$
at $F \geq 50$ mJy, quite independently of the frequency.

\item We expect an induced fluctuation amplitude of the CMB map of
order of $C_{\ell,Blazar} \approx 31 Jy^2/sr$ which corresponds to a temperature
fluctuation of $\sim 0.013 \mu K^2 sr$ at 41 GHz. The corresponding Poissonian
spectrum of the Blazar population equals the CMB power spectrum at $\ell \approx
1000$, and contaminates it at levels $\sim 20\%$ at $\ell \approx 500$ and $\sim
50\%$ at $\ell\approx 800$.

\item The previous estimates of $C_{\ell,Blazar}$ are a firm lower limit to the
level of CMB fluctuation induced by the Blazar population. In fact, the value of
the Poissonian term $C_{\ell,Blazar}$ is likely to be higher by a factor at
least $\simgt 2-3$ considering both the Blazar variability  and the contribution
of steep-spectrum sources at low frequencies, radio-galaxies and possibly
starburst galaxies at lower fluxes. We also stress that allowance for an
appropriate clustering term might enhance substantially the angular power
spectrum of the point-like sources and hence increase the level of contamination
of the CMB power spectrum.

\item It is important to stress that flux variability at millimeter wavelengths may
be very substantial (higher than the factors 3-10 on time scales of weeks to
months seen at cm wavelengths) and definitely increases the level of
contamination of CMB maps when these are built over long integration periods.
Moreover, the Blazar variability increases with increasing frequency so that
such a bias is more relevant for high-frequency CMB detections.

\item As a consequence of the non-negligible Blazar contamination of the CMB
maps, the primordial CMB polarization seems to be greatly affected by the
presence of polarized Blazars.
Thus, the significant contamination of CMB fluctuation maps and the
non-negligible fraction of residual unresolved sky brightness will render
difficult to detect primordial polarization fluctuations.

\end{itemize}

The many high-sensitivity and high-resolution CMB experiments planned for the
near future will definitely face with the problems highlighted in this paper. In
fact, we predict at least 1300 sources/sr at 90 GHz which produce a flux excess
corresponding to $>10 ~\mu K$ in a pixel of 15 arcmin size.
One proven way of finding Blazars down to faint radio fluxes ($\sim $ mJy) is to
use highly efficient multi-frequency selection techniques, similarly to the
DXRBS survey which has an efficiency of $\sim 90 \%$ (\cite{lan01}), by cross
correlating deep radio surveys (e.g., NVSS, FIRST etc.) with deep X-Ray data
over the largest possible area of sky. However, to reach flux levels lower than
$\sim 10$ mJy, this approach requires an X-ray sensitivity of $\sim
10^{-15}$\ergs. Such a sensitivity is reached in XMM-Newton and Chandra X-ray
images only for a very limited area of sky.
In conclusion, it is crucial to determine the level of Blazar contamination in
precision CMB experiments in order to separate the contaminating foreground from
the primordial CMB fluctuations and to provide a reliable determination of the
cosmological parameters.

\begin{acknowledgements}

The authors are grateful to E. Massaro for useful discussions, to P. Padovani
for providing DBRBS results in advance of publication, and to M. Perri, E.
Cavazzuti and S. Piranomonte for their help in the analysis of part of the
NVSS-RASS 1JY data. We also thank N. Panagia for useful comments on a
preliminary version of the paper. This research has made use of data from the
ASI Science Data Center and from the NASA/IPAC Extragalactic Database (NED).

\end{acknowledgements}


\begin{thebibliography}{}
\bibitem[Bade et al. 1998]{bad98}  Bade N. et al. Astron. Astrophys. 334, 450
\bibitem[Bennett et al. 2003]{ben03a} Bennett C.L., Bay M., Halpern M. et al. 2003a ApJ, 583, 1
\bibitem[Bennett et al. 2003]{ben03b} Bennett C.L., 2003b ApJ submitted, astro-ph/0302208
\bibitem[Blandford \& Rees 1978]{blarees} Blandford R.D. \& Rees M.J. 1978
in Pittsburgh Conf. on BL Lac objects, p. 328
\bibitem{} Bond, J., Efstathiou, G. and Tegmark, M. 1997, MNRAS, 291, 33
\bibitem[Coble et al. (2003)]{cob03} Coble K., et al. 2003, ApJS, submitted, astro-ph/0301599
\bibitem[De Bernardis et al. 2000]{deber00}De Bernardis P. et al. 2000, {\it Nature}, 404, 955
\bibitem[Hasinger 2000]{has00}Hasinger G., in ISO Surveys of a Dusty Universe,
D. Lemke, M. Stickel, K. Wilke eds., Springer, astro-ph/0001360
\bibitem[Giommi et al. 1999]{gio99} Giommi P., Menna M. T., Padovani P., 1999, MNRAS, 310, 465
\bibitem[Giommi et al. (2002a)]{gio02a} Giommi P., et al. 2002a, Proceeding of the
workshop "Blazar Astrophysics with BeppoSAX and Other Observatories", Giommi,
Massaro \& Palumbo eds., ASI special publication, p. 63
\bibitem[Giommi et al. 2002b]{gio02b} Giommi P., Perri M., Landt H., Perlman E.
2002b, Proceeding of the workshop "Blazar Astrophysics with BeppoSAX and Other
Observatories", Giommi, Massaro \& Palumbo eds., ASI special publication, p. 133
\bibitem[Giommi et al. 2002c]{gio02c} Giommi P., Perri M., Piranomonte S.,
Padovani P. 2002c, Proceeding of the workshop "Blazar Astrophysics with BeppoSAX
and Other Observatories", Giommi, Massaro \& Palumbo eds., ASI special
publication, p. 123
\bibitem{Grainge et al. 2002} Grainge, K. et al. 2002, MNRAS in press
(astro-ph/0212495)
\bibitem{Halverson et al. 2001} Halverson, N.W. et al. 2002, ApJ, 568, 38
\bibitem{Hanany et al. 2000} Hanany, S. et al. 2000, ApJ, 545, L5
\bibitem[Hu & White 1997]{huwhite97} Hu, W. \& White, M. 1997, New Astronomy, 2, 323
\bibitem[Hughes et al. 1992]{hughes92} Hughes, P. A., Aller, H. D. \& Aller M.F.,  1992 ApJ, 396, 469
\bibitem{} Jungman, G., Kamionkowski, M., Kosowski, A. and Spergel, D. 1996,
Pys.Rev.D, 54, 1332
\bibitem[Landt et al. 2001]{lan01} Landt H., Padovani P., Perlman E.S., Giommi P.,
Bignall H. and Tzioumis A., 2001, M.N.R.A.S. 323, 757
\bibitem[Laurent-Muehleisen et al. 1998]{mue98}
Laurent-Muehleisen S. A., Kollgaard R. I., Ciardullo R., Feigelson E. D., et
al., 1998, ApJS, 118, 127
\bibitem{} Myers, S.T. et al. 2002, preprint astro-ph/0205385
\bibitem[Padovani 2002]{pad02} Padovani P., 2002, Proceeding of the workshop
"Blazar Astrophysics with BeppoSAX and Other Observatories, Giommi, Massaro \&
Palumbo eds., ASI special publication, p. 101.
\bibitem{Pearson et al. 2002} Pearson, T.J. et al. 2002, ApJ in press
(astro-ph/0205388)
\bibitem[Peebles (1980)]{Peeb80} Peebles, P.J.E. 1980, The Large-Scale Structure
of the Universe (Princeton Univ. Press, Princeton)
\bibitem[Perri et al. (2002)]{perri02} Perri M., Giommi P., Piranomonte S., Padovani P.,
2002, Proceeding of the workshop "Blazar Astrophysics with BeppoSAX and Other Observatories", Giommi,
Massaro \& Palumbo eds., ASI special publication, p. 119
\bibitem{} Pierpaoli, E. 2003, ApJ, 589, 58
\bibitem[Rector et al. 2000]{rec00} Rector T.A., et al. 2000, Astron. J. 120, 1626
\bibitem[Urry \& Padovani 1995]{urrpad95} Urry C.M, Padovani P., 1995 Publ. Astr. Soc. Pacific 107,803
\bibitem[Padovani et al. 2003]{pad03}Padovani P., et al. 2003, in preparation
\bibitem{} Refregier, A. \& Spergel, D.N. 2000, ApJ, 531, 31
\bibitem[Scott \& White (1999)]{sw99} Scott, D. and White, M. 1999, A\&A, 346, 1
\bibitem{} Smooth, G.F. et al. 1992, ApJ, 396, L1
\bibitem[Stickel et al. 1991]{sti91}Stickel M., Padovani P., Urry C.M. Fried J.W.,
Ku\"hr H., 1991, ApJ 374,431
\bibitem{taylor03} Taylor, A. et al. 2003, MNRAS in press (astro-ph/0205381)
\bibitem[Tegmark \& Efstathiou (1996)]{tf96} Tegmark, M. and Efstathiou, G.
1996, MNRAS, 281, 1297
\bibitem[Toffolatti et al. (1998)]{tof98} Toffolatti L., et al. 1998, MNRAS, 297, 117
\bibitem[Wall \& Peacock 1985]{wp85} Wall J.V., Peacock J.A.,  M.N.R.A.S. 216, 173

\end{thebibliography}
\end{document}